\documentclass[10pt,twoside]{epnt1p}
\usepackage{graphics}
\usepackage{graphicx}
\usepackage{epsfig}

\usepackage{mcite}

\usepackage{amssymb}
\usepackage{amsmath}
\usepackage{url}
\newcommand{\E}{\rm EHEC\nu}
\newcommand{\C}{\rm {C}\nu{\rm B}}
\begin{document}
{\hfill\normalsize \tt DESY 06-210}

\begin{frontmatter}


\title{Probing the variation of relic neutrino masses with extremely high-energy cosmic neutrinos}


\author[address1]{Lily~Schrempp},

\address[address1]{Deutsches Elektronen-Synchrotron DESY, Notkestra\ss e  85, 22607 Hamburg, Germany}
\ead{lily.schrempp@desy.de}

\begin{abstract}

We analyze the prospects for testing the cosmic neutrino background and its interpretation as source of Neutrino Dark Energy with the radio telescope LOFAR.  

\end{abstract}



\end{frontmatter}

\section{\label{sec:introd} Introduction}

Certainly, one of the most challenging questions in modern Cosmology and Particle physics is, what is the nature of Dark Energy? Strong observational evidence hints at the existence of this smooth, exotic energy component which drives the apparent accelerated expansion of the universe. Recently, Fardon, Nelson and Weiner~\cite{Fardon:2003eh,Hung:2000yg,Gu:2003er} have shown that Big Bang relic neutrinos -- the analog of the cosmic microwave photons -- are promoted to a natural Dark Energy candidate if they interact through a new non-Standard Model force. Due to this scalar force, the homogeneously distributed relic neutrinos can form a negative pressure fluid and thus exhibit just the right properties to act as Dark Energy. As a further consequence of this new interaction, the neutrino mass becomes a function of neutrino energy density which decreases as the universe expands. Thus, intriguingly, the neutrino mass is not a constant but is promoted to a dynamical quantity. After discussing the details of this so-called Mass Varying Neutrino (MaVaN) scenario, I will consider an astrophysical possibility of testing it with extremely high energy-cosmic neutrinos ($\E$). A more detailed discussion of the results may be found in Ref.~\cite{Ringwald:2006ks}.   
 
\section{\label{sec:MaVaNs} Mass Varying Neutrinos}

I will concentrate on a concrete realization of the non-Standard Model neutrino interaction as preferred in the literature~\cite{Fardon:2003eh,Fardon:2005wc,Spitzer:2006hm} which implements the seesaw mechanism~\cite{Gell-Mann,Yanagida,Minkowski,Mohapatra:1979ia}. Generically, varying mass particle scenarios exhibit a so-called dark sector which only consists of Standard Model singlets. In the considered case of MaVaNs it contains a light scalar field, the acceleron ${\mathcal{A}}$, which has an associated fundamental potential $V_0({A})$. The acceleron interacts with a second field of the dark sector, a right-handed neutrino $N$, through a Yukawa coupling, $\kappa{\mathcal{A}}NN$, and thus generates its mass $M_N({\mathcal{A}})=\kappa {\mathcal{A}}$. The dark force mediated by the acceleron is transmitted to the active neutrino sector via the seesaw mechanism which in addition provides a natural explanation for the smallness of the left-handed neutrino mass $m_\nu$. Accordingly, at scales well below $100$ GeV the Lagrangian contains a Majorana mass term~\cite{Fardon:2003eh,Ringwald:2006ks},
\begin{equation}
{\mathcal{L}}\supset \frac{m_D^2}{M_N({\mathcal{A}})}\nu^2+h.c.+V_0({\mathcal{A}}),\,\,\mbox{where}\,\,m_\nu(\mathcal{A})= \frac{m_D^2}{M_N({\mathcal{A}})},
\end{equation}
for the active left-handed neutrino, where the Dirac type mass $m_D$ originates from electroweak symmetry breaking and $M_N\gg m_D$ has been assumed. Consequently, the neutrino mass $m_\nu(\mathcal{A})$ is light and, since it is generated by the value of the acceleron, neutrinos interact through a new force mediated by ${\mathcal{A}}$.

\begin{figure*}[t]
\begin{minipage}[t]{0.48\linewidth}
\centering\epsfig{file=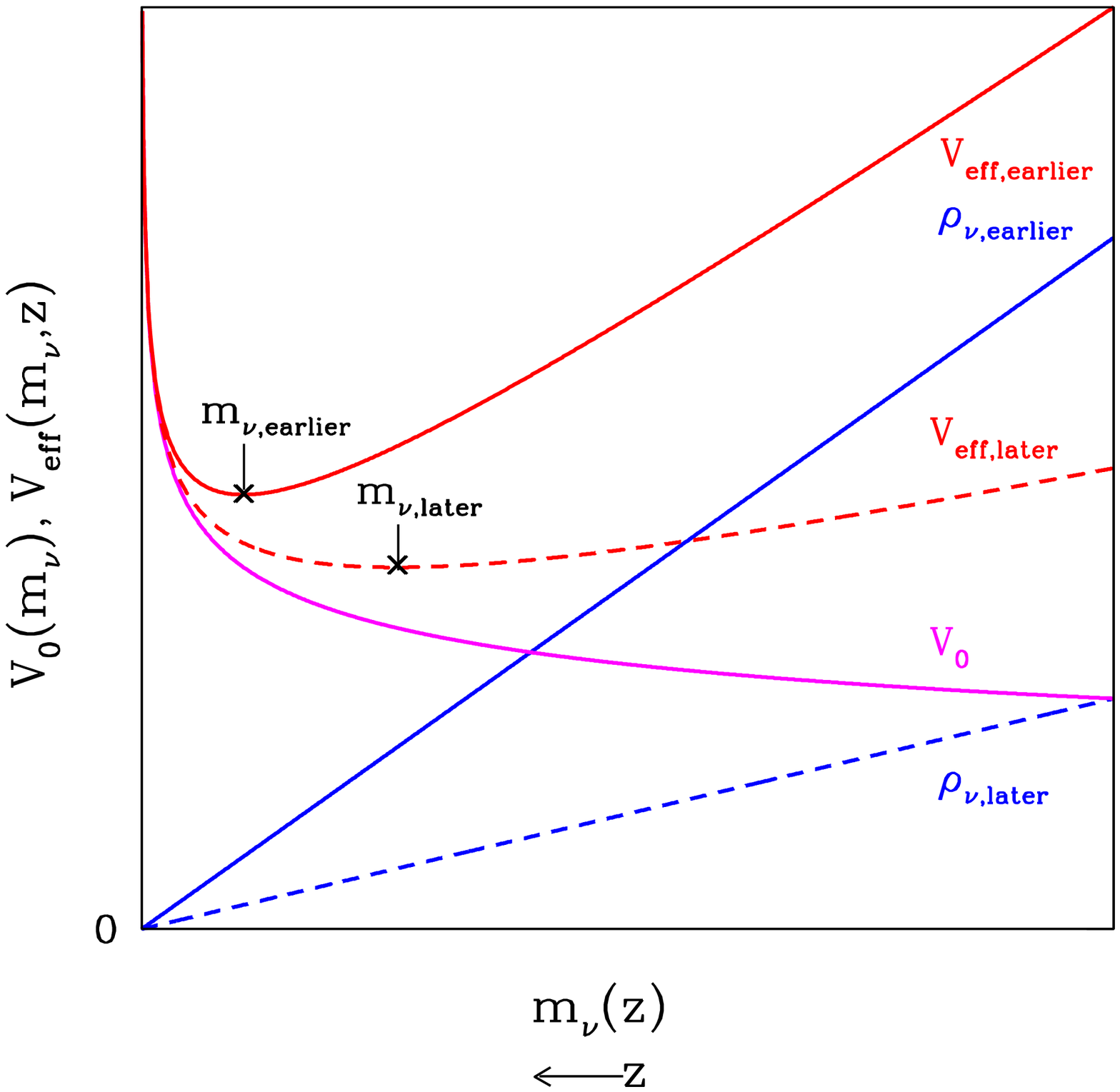,width=\linewidth}
\end{minipage}\hfill
\begin{minipage}[t]{0.48\linewidth}
\centering\epsfig{file=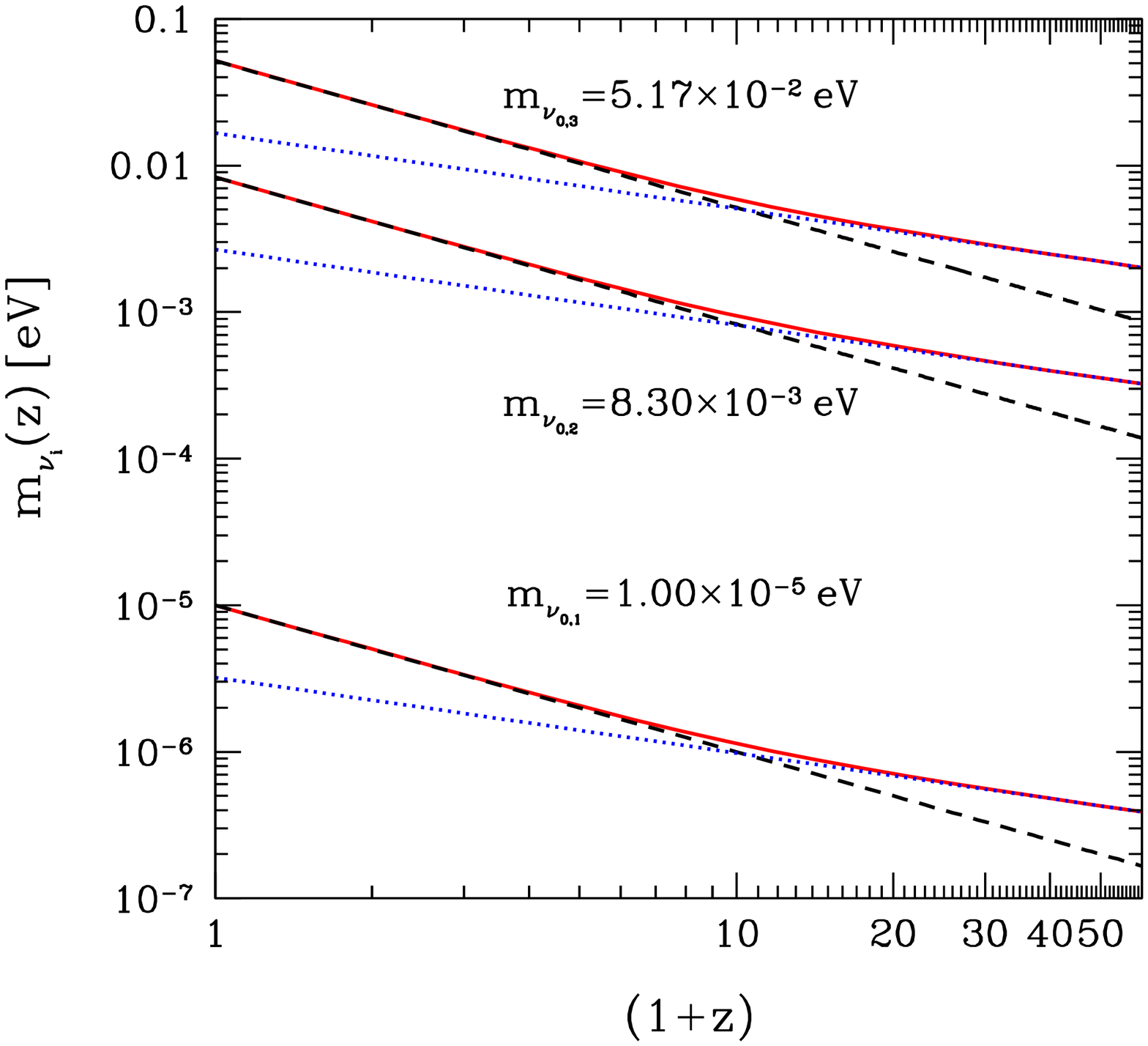,width=\linewidth}
\end{minipage}
\caption[]{a) Evolution of the effective potential $V_{\rm eff}(m_\nu,z)$ and the neutrino mass $m_\nu(z)$ due to changes in the neutrino energy density $\rho_\nu(z)$; b) Exact $m_{\nu_i}(z)$ (solid), approximated by $m_{\nu_i}(z)\propto (1+z)^{-1}$ and $m_{\nu_i}(z)\propto (1+z)^{-1/2}$ (dashed and dotted) in the low and high redshift regime, respectively, assuming $m_{\nu_{0,1}}=10^{-5}\,{\rm eV}$ and a normal neutrino mass hierarchy such that neutrino oscillation experiments fix $m_{\nu_{0,2}}=8.3\times 10^{-3}\,{\rm eV}$ and $m_{\nu_{0,3}}=5.17\times 10^{-2}\,{\rm eV}$ today (fig.~\ref{Mass}b) from Ref.~\cite{Ringwald:2006ks}).}\label{Mass}
\end{figure*}

The coupling leads to a complex interplay between the acceleron and the neutrinos which links their dynamics. Since the neutrino energy density is a function of the neutrino mass $m_\nu({\mathcal{A}})$, it becomes an indirect function of the value of the acceleron, $\rho_\nu(m_\nu({\mathcal{A}}),z)$. As a direct consequence, it stabilizes the acceleron by contributing to its effective potential,
\begin{eqnarray}
V_{\rm eff}({\mathcal{A}},z)&=&\rho_\nu(m_\nu({\mathcal{A}}),z)+V_0({\mathcal{A}}),\,\,\mbox{where}\\
\rho_{\nu}(m_\nu({\mathcal{A}}),z)&=&\frac{T_{\nu}(z)^4}{\pi^2} \int\limits_0^{\infty}\frac{dy\, y^2 \sqrt{y^2+\left(\frac{m_{\nu}({\mathcal{A}})}{T_{\nu}(z)}\right)^2}}{e^y+1},
\end{eqnarray}
with $z$ denoting the redshift and $T_{\nu_0}(z)=T_{\nu_0}(1+z)$ the neutrino temperature with $T_{\nu_0}\sim 1.69\times 10^{-4}$ eV. Since cosmic expansion causes a dilution of the neutrino energy density, also $V_{\rm eff}$ evolves with time. Assuming the curvature scale of $V_{\rm eff}$ and thus the mass of $\mathcal{A}$ to be much larger than the Hubble scale, $\partial V^2_{\rm eff}/\partial{\mathcal{A}}^2= m^2_{\mathcal{A}}\gg H^2$, the adiabatic solution to the equations of motions apply~\cite{Fardon:2003eh}. As a result, the acceleron instantaneously tracks the minimum of its effective potential, the total energy density of the coupled system, and thus varies on cosmological time scales. Consequently, the neutrino mass $m_\nu(\mathcal{A})$ which is generated from the value of the acceleron, is not a constant but promoted to a dynamical quantity. As long as $\partial m_\nu/\partial {A}$ does not vanish, its time variation is determined by 
\begin{equation}
\label{minimum}
\frac{\partial V_{\rm eff}(\mathcal{A})}{\partial\mathcal{A}}=\frac{\partial m_\nu(\mathcal{A})}{\partial\mathcal{A}}\left(\left.\frac{\partial \rho_\nu(m_\nu,z)}{\partial  m_\nu}\right|_{m_\nu=m_\nu(\mathcal{A})}+\left.\frac{\partial V_0(m_\nu)}{\partial m_\nu}\right|_{m_\nu=m_\nu(\mathcal{A})}\right)=0.
\end{equation} 
Fig.~\ref{Mass}a) shows the evolution of both the effective potential $V_{\rm eff}(\mathcal{A},z)$ as well as its minimum due to changes in $\rho_\nu(z)$. 

In the following I will address a possible shortcoming of MaVaN models and how it can be avoided. Since MaVaNs attract each other through the force mediated by the acceleron, they are possibly subject to a phenomenon similar to gravitational instabilities of cold dark matter (CDM)~\cite{Afshordi:2005ym}. As long as the neutrinos are still relativistic, their random motions can prevent a collapse. However, when they turn non-relativistic the system can become unstable leading to the possible formation of so-called `neutrino nuggets'~\cite{Afshordi:2005ym}. If the neutrinos really clump, they would redshift like CDM with $w\sim 0\neq -1$ and thus cease to act as Dark Energy. However, firstly, certain constraints on the scalar potential $V_0$ and on the function $m_\nu({A})$, can lead to stable MaVaN models even in the highly non-relativistic regime~\cite{Takahashi:2006jt,Bjaelde2006}. Secondly, neutrino oscillation experiments allow one neutrino to be still relativistic today, which could be responsible for cosmic acceleration until the present time~\cite{Afshordi:2005ym,Fardon:2005wc,Spitzer:2006hm}. In Ref.~\cite{Fardon:2005wc} the latter case emerges naturally after a straightforward super-symmetrization of the standard MaVaN scenario. However, the modified model relies on a slightly different acceleration mechanism known from Hybrid inflation~\cite{Linde:1993cn} to be considered in the following. A scalar field keeps a second scalar field in a metastable minimum due to its large value and the energy density stored in the false minimum can drive cosmic acceleration. In the MaVaN hybrid model the first scalar field can be identified with the acceleron, which is driven to larger values due to the stabilizing effect of the fermionic neutrino background. It keeps the scalar partner ${N}$ of the lightest neutrino, naturally present in a super-symmetric theory, in a false minimum until it reaches a critical value ${\mathcal{A}}_{\rm crit}$. The combined scalar potential $V_0({\mathcal{A}},{N})$ appears as dark energy with $w\sim -1$ and can therefore drive accelerated expansion. This hybrid model provides a microscopic origin for a quadratic scalar potential $V_0({\mathcal{A}})\propto {\mathcal{A}}^2$ and thus according to eq.~\ref{minimum} fixes the neutrino mass evolution. After generalizing eq.~\ref{minimum} to include three neutrino species, the evolution of the neutrino masses in the low redshift regime is found to be well approximated by a simple power law~\cite{Ringwald:2006ks},
\begin{equation}
\label{mvary}
m_{\nu_i}(z)\simeq m_{\nu_i,0}(1+z)^{-1},\,\,\mbox{where}\,\,m_{\nu_i,0}=m_{\nu_i}(0)\,\,\mbox{and}\,\,i=1,2,3,
\end{equation}
as shown in fig.~\ref{Mass}b), where the lightest neutrino is assumed to be still relativistic today, $m_{\nu_1}=10^{-5}\,{\rm eV}\ll T_{\nu_0}$. Finally, we are in a position to do MaVaN phenomenology. In the following I will discuss a possible astrophysical test~\cite{Ringwald:2006ks} for the neutrino mass variation involving extremely high-energy cosmic neutrinos.

\section{\label{sec:EHECnu} Extremely high-energy cosmic neutrinos} 

\begin{figure*}[t]
\begin{minipage}[t]{0.48\linewidth}
\centering\epsfig{file=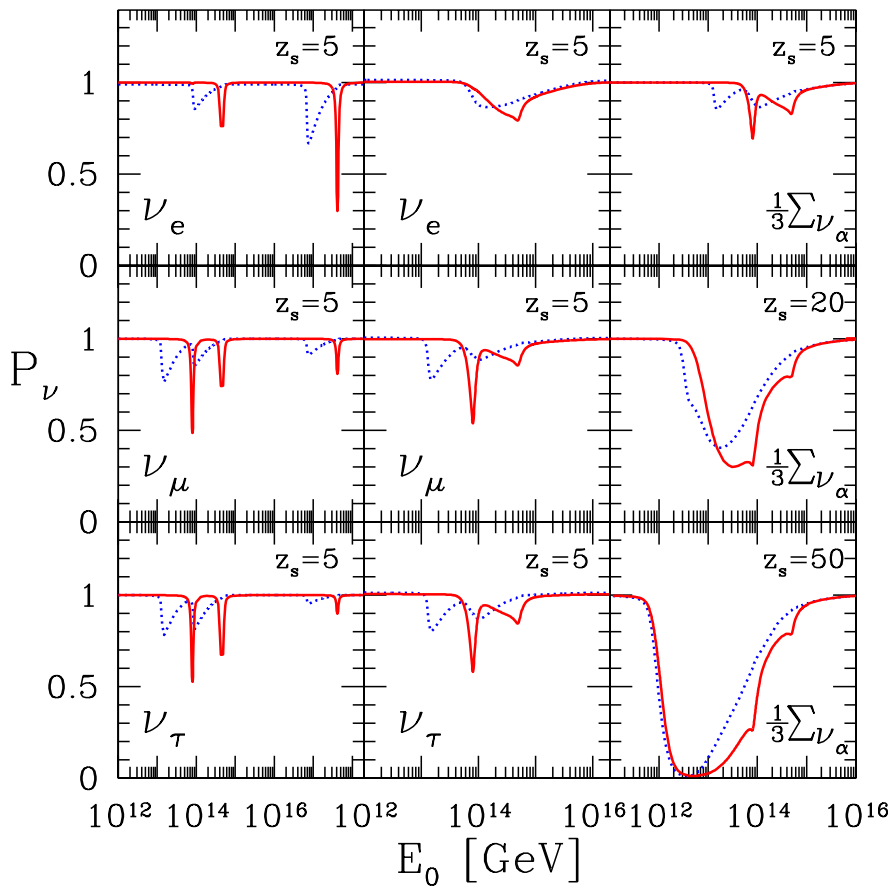,width=11cm}
\end{minipage}\hfill
\caption[]{a), b): Flavor survival probability $P_\nu=P_{\nu_{\alpha}}$, $\alpha=e,\mu,\tau$, without and with inclusion of thermal background effects, respectively, integrated back to $z_s=5$; c) Normalized sum $P_\nu=\frac{1}{3}\sum P_{\nu_{\alpha}}$ including thermal effects for $z_{\rm s}=5$, $z_{\rm s}=20$ and $z_{\rm s}=50$ from top to bottom. All panels assume a normal neutrino mass hierarchy with $m_{\nu_{0,1}}=10^{-5}$ eV and varying (solid) and constant (dotted) neutrino masses as a function of their energy $E_0$ at earth (figures from Ref.~\cite{Ringwald:2006ks}).}\label{Dips}
\end{figure*}

We are living in exciting times for extremely high-energy cosmic neutrinos~\cite{Ringwald:2005wa,Ringwald:2006ks}. Existing and planned observatories cover an energy range of $10^7\,{\rm GeV}<E_0<10^{17}$ GeV and promise appreciable event samples (cf.~\cite{Ringwald:2005wa,Ringwald:2006ks} and references therein). Thus it seems timely to consider the diagnostic potential of $\E$'s for astrophysical processes. A particular example will be discussed in the following. If the energy of an $\E$ coincides with the resonance energy, $E_{i}^{\rm res}=\frac{M^2_Z}{2m_{\nu_{i}}}=4.2\times 10^{12}\,\,\left(\frac{\rm eV}{m_{\nu_i}}\right) {\rm GeV},\,i=1,2,3,$ of the process $\nu\bar{\nu}\rightarrow Z$, 
an $\E$ can annihilate with a relic anti-neutrino and vice versa into a $Z$ Boson~\cite{Weiler:1982qy,Weiler:1983xx,Gondolo:1991rn,Roulet:1992pz,Yoshida:1996ie,Eberle:2004ua,Barenboim:2004di,D'Olivo:2005uh} with mass $M_Z$. This exceptional loss of transparency of the cosmic neutrino background ($\C$) with respect to cosmic neutrinos is expected to lead to sizeable absorption dips in the diffuse $\E$ fluxes to be detected at earth (cf. fig.~\ref{Dips}). Independent of the nature of neutrino masses, their resolution would constitute the most direct evidence of the $\C$ so far. 
In addition, since the annihilation process is sensitive to the neutrino mass and thus to its possible time variation, it could serve as a test for Neutrino Dark Energy (MaVaNs). 
In fig.~\ref{Dips} the survival probability of an $\E$ both for varying and constant neutrino masses is plotted as a function of its energy $E_0$ as measured at earth. It encodes the physical information on possible annihilation processes (for details see~\cite{Ringwald:2006ks,Weiler:1982qy,Weiler:1983xx,Gondolo:1991rn,Roulet:1992pz,Yoshida:1996ie,Eberle:2004ua,Barenboim:2004di,D'Olivo:2005uh}) and takes values between $0$ and $1$. In order to disentangle the different influences on the $\E$ survival probability, let us first assume the relic neutrinos to be at rest (cf. fig.~\ref{Dips}a)). In this case the absorption features for constant neutrino masses are only subject to the effect of cosmic redshift which causes an $\E$ emitted at its source at $z_{\rm s}$ with energy $E$ to arrive at earth with the redshifted energy $E_0=E/(1+z_{\rm s})<E$. As a result, the survival probability is reduced in the interval $E_{0,i}^{\rm res}/(1+z_{\rm s})<E_0<E_{0,i}^{\rm res}$ and the absorption minima are thus redshift distorted. In case of a possible mass variation, however, the survival probability exhibits sharp spikes at the resonance energies $E_{0,i}^{\rm res}$ (cf. fig.~\ref{Dips}a)). This can be understood by recalling the mass behavior in the low redshift regime stated in eq.~\ref{mvary}. The neutrino masses $m_{\nu_i}(z)$ introduce a $z$ dependence into the resonance energies, $E_{i}^{\rm res}(z)=M^2_Z (1+z)/(2m_{\nu_{0,i}})=E_{0,i}^{\rm res}(1+z)$ which is compensated by the effect of cosmic redshift. Accordingly, all annihilations occurring at $0<z<z_{\rm s}$ contribute to an absorption dip at $E_{i}^{\rm res}(z)/(1+z)=E_{0,i}^{\rm res}$. As a second step let us take into account that the relic neutrinos are moving targets with a Fermi-Dirac momentum distribution. In fig.~\ref{Dips}b) the corresponding Fermi-smearing both for constant and varying neutrino masses results in a thermal broadening (and thus merging) of the dips produced by the mass eigenstates (cf. fig.~\ref{Dips}a)). 

The plot of the flavor summed survival probability in fig.~\ref{Dips}c) shows that the dip depth increases with $z_s$ independent of the nature of neutrino masses. However, generically, the MaVaN features are clearly shifted to higher energies and the minima deeper than for neutrinos with constant mass.   
\section{\label{sec:outlook} Outlook} 
\begin{figure*}[t]
\begin{minipage}[t]{0.48\linewidth}
\centering\epsfig{file=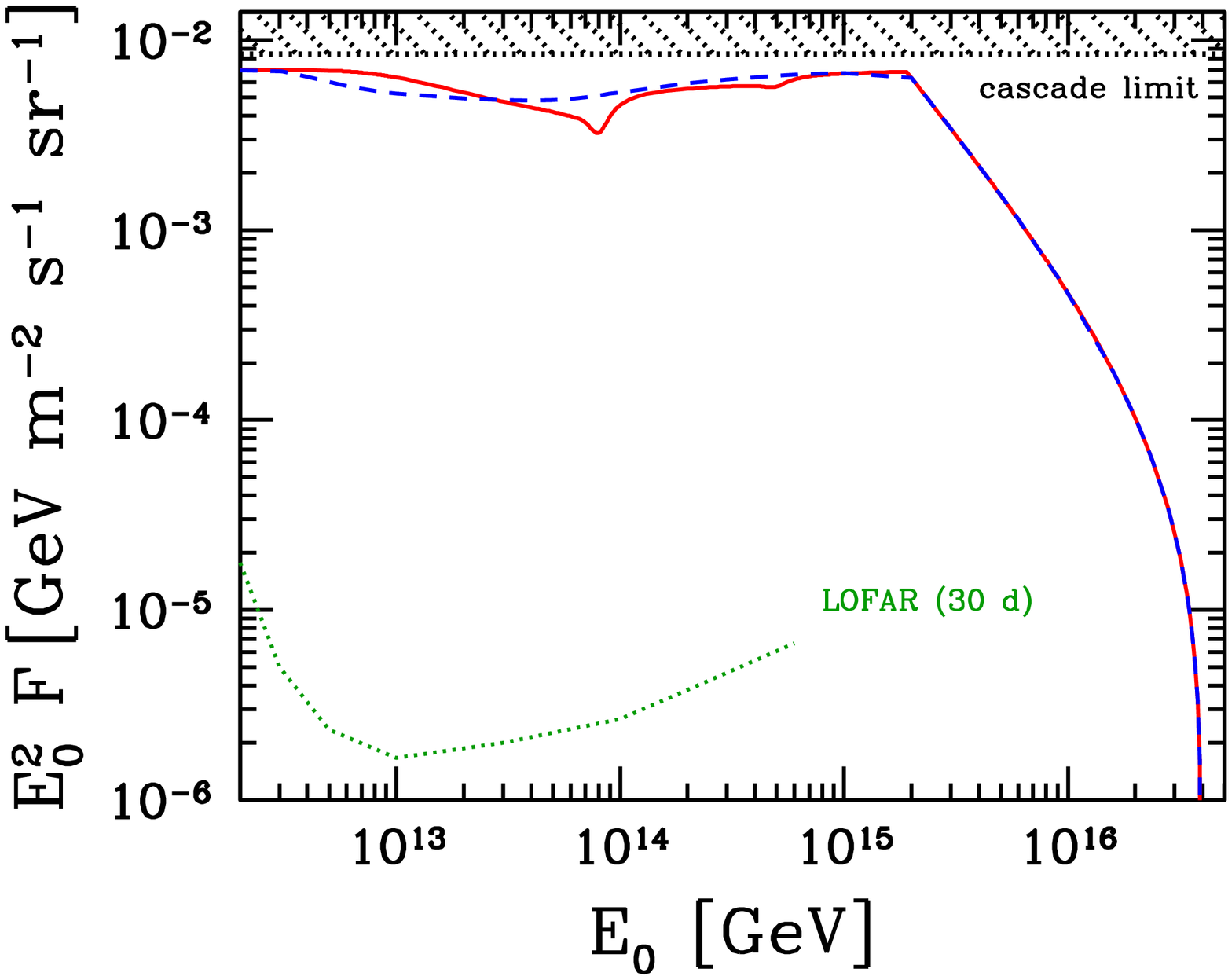,width=\linewidth}
\end{minipage}\hfill
\begin{minipage}[t]{0.48\linewidth}
\centering\epsfig{file=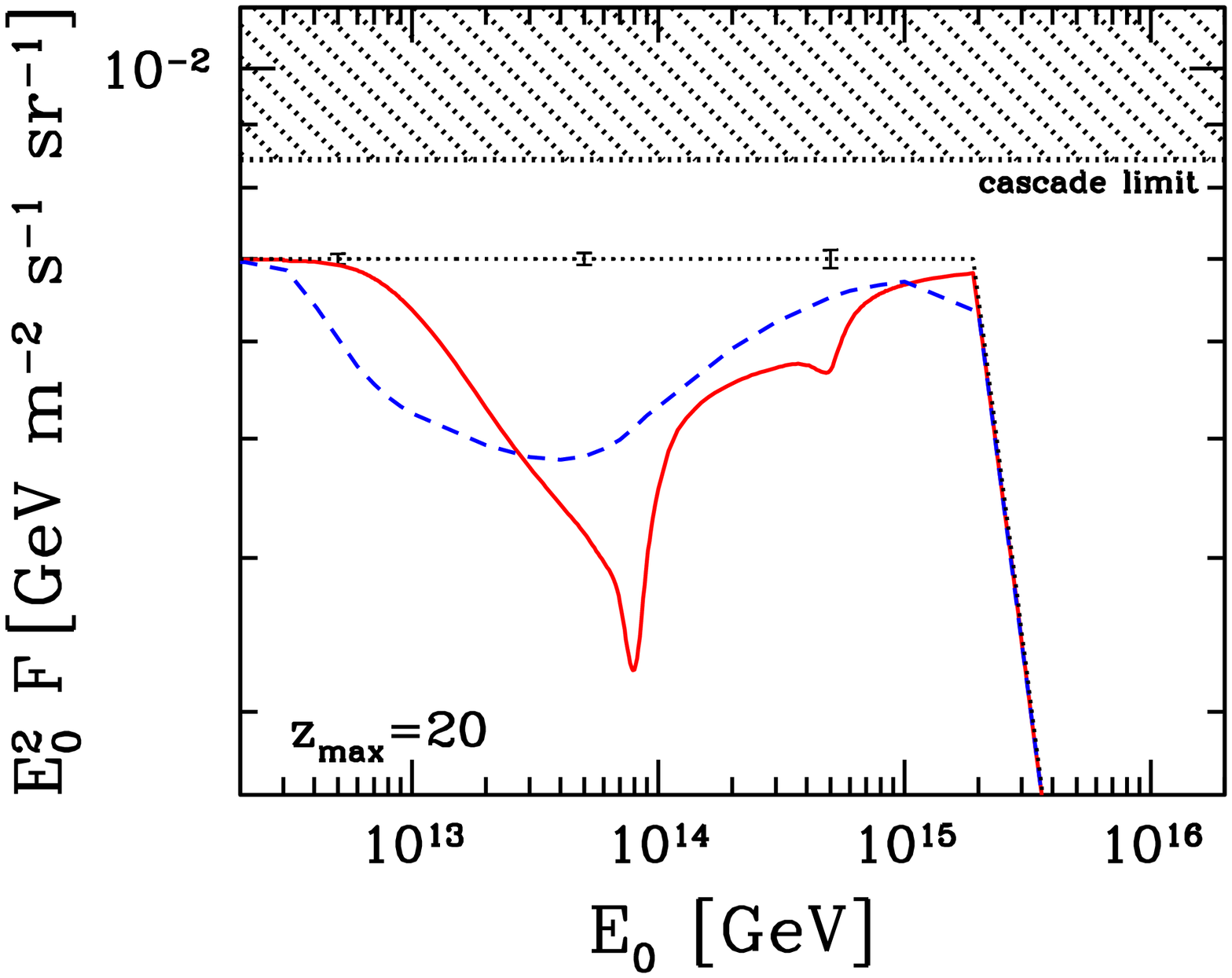,width=\linewidth}
\end{minipage}
\caption[]{$E^2_0 F$ with $F=\sum F_{\nu_{\alpha}}+\sum F_{\bar{\nu}_{\alpha}}$ for a flux saturating today's observational bound for varying (solid) and constant (dashed) neutrino masses and for $z_{\rm s}=20$, assuming a normal neutrino mass hierarchy with $m_{\nu_{0,1}}=10^{-5}$ eV. a) Together with projected sensitivity of LOFAR expressed in terms of the diffuse neutrino flux per flavor, corresponding to one event per energy decade and indicated duration; b) Together with error bars indicating the corresponding statistical accuracy (figures from Ref.~\cite{Ringwald:2006ks}).}\label{Astro}
\end{figure*}
The fig.~\ref{Astro} shows the expected $\E$ flux from an astrophysical source at a redshift $z_{\rm s}=20$ both for constant was well as for varying neutrino masses normalized to saturate today's observational bound. In fig.~\ref{Astro}a) it is plotted together with the projected sensitivity of the radio telescope LOFAR~\cite{Scholten:2005pp} which corresponds to maximally $3500$ events detected per energy decade and indicated duration. This translates into very small statistical error bars as included in fig,~\ref{Astro}b), a blow up of fig.~\ref{Astro}a). Accordingly, independent of the nature of neutrino masses, the detection of absorption dips with LOFAR and thus the most direct evidence for the existence of the $\C$ so far can be expected within the next decade. Furthermore, if LOFAR achieves a decent energy resolution, the variation of neutrino masses and thus the interpretation of the $\C$ as source of Dark Energy could be tested.  

\section*{\bf Acknowledgments}

{I thank Andreas Ringwald for helpful discussions and encouragement, Markus Ahlers and Yvonne Wong for technical advice and Yvonne Wong for discussions.
}

\frenchspacing
\bibliography{schrempp}
\addcontentsline{toc}{section}{Bibliographie}

\bibliographystyle{utcaps}

\setcounter{section}{0}
\setcounter{subsection}{0}
\setcounter{figure}{0}
\setcounter{table}{0}
\newpage
\end{document}